\documentclass[english,twocolumn,preprintnumbers,superscriptaddress,letterpaper]{revtex4}
\usepackage{epsfig,psfrag}
\usepackage{amssymb,amsmath}
\usepackage{hyperref}
\usepackage{graphicx}
\newcommand{\beq}{\begin{equation}}
\newcommand{\eeq}{\end{equation}}
\newcommand{\beqa}{\begin{eqnarray}}
\newcommand{\eeqa}{\end{eqnarray}}
\newcommand{\Op}{\mathcal{O}}
\newcommand{\superN}{\mathcal{N}}
\newcommand{\DK}{\Delta_\mathcal{K}}

\catcode`\@=11

\begin{document}
\preprint{Imperial/TP/09/AR/01}
\title{Konishi at strong coupling from ABE}
\author{Adam Rej$^{1}$, Fabian Spill}
\email{adam.rej, fabian.spill@imperial.ac.uk}
\affiliation{Blackett Laboratory, Imperial College, London SW7 2AZ, U.K.}

\begin{abstract}
In this letter, we derive analytically the scaling dimension of the Konishi operator in planar $\superN=4$ gauge theory at strong coupling from the asymptotic Bethe equations. The first two leading terms agree with the recent string computation and numerical analysis of TBA equations. At the third order we find a spurious logarithm of the coupling constant,  which should be \textit{absent} in the anomalous dimension of any finite operator of planar $\superN=4$ SYM theory. Showing the cancelation of this term would provide an important test at strong coupling for the recently proposed sets of TBA equations for the planar AdS/CFT correspondence. 

\end{abstract}

\maketitle

\section{Introduction and Summary}
\label{sec:intro}

The planar $\superN=4$ gauge theory exhibits a very remarkable property absent in most gauge theories. Namely, the dilatation operator is believed to be asymptotically integrable. In the groundbreaking paper \cite{Minahan:2002ve}, the one-loop integrability of the dilatation operator in certain subsectors of the gauge side of the AdS/CFT correspondence was discovered. Later on \cite{Beisert:2003jj} the complete one-loop dilatation operator has been found and the corresponding one-loop Bethe equations were written down \cite{Beisert:2003yb}. After many non-trivial steps \cite{fussnote}, the form of the all-loop asymptotic Bethe equations (ABE) was conjectured \cite{Beisert:2005fw} up to the so called dressing factor \cite{Arutyunov:2004vx}, which only contributes starting from the four-loop order. Subsequently, relying on the crossing equation proposed in \cite{Janik:2006dc} and assuming certain transcendentality properties, it was possible to uniquely fix this factor \cite{Beisert:2006ez}. In this way the asymptotic spectrum of the planar $\mathcal{N}=4$ SYM theory has been completely determined. The asympoticity of these equations means that for a generic operator with $L$ constituent fields the corresponding anomalous dimension can be calculated correctly up to the $\mathcal{O}(g^{2L})$ order.

Beyond this order the asymptotic Bethe equations are \textit{not} valid, as it was shown in \cite{Kotikov:2007cy}. The asymptotic integrability was established by disregarding the contribution of certain class of Feynman diagrams, while studying the structure of mixing of the operators. These diagrams are commonly referred to as wrapping diagrams, the reason being their topological properties \cite{Sieg:2005kd}. In the spin chain picture, these diagrams correspond to the interactions wrapping around the spin chain. Since the interaction between two nearest neighbours provides a factor of $g^2$, the first of the wrapping diagrams may appear at the order $\Op(g^{2 L})$, where $L$ is the length of the spin chain. For twist-two (length-two) operators the supersymmetry \textit{delays} the wrapping interactions to the four-loop order. 

In the seminal paper of \cite{Bajnok:2008bm} it was proposed to evaluate the first L\"uscher corrections, previously studied for the $AdS_5 \times S^5$ sigma model and single excitation in \cite{Janik:2007wt}, to calculate the leading wrapping corrections to the shortest unprotected operator in the $\mathfrak{sl}(2)$ sector, i.e. Konishi operator. The result remarkably coincides with very complicated field theory computations \cite{Fiamberti:2007rj,Velizhanin:2008jd}.

As for other integrable sigma models, the full solution of the spectral problem is believed to be given in terms of a set of TBA equations. There has been recently a lot of progress in this direction \cite{Ambjorn:2005wa}-\cite{Arutyunov:2009zu} and ultimately the Y-system \cite{Gromov:2009tv} and the TBA equations for the ground state  \cite{Bombardelli:2009ns}-\cite{Arutyunov:2009ur} have been formulated. The authors of \cite{Gromov:2009bc} have also proposed the TBA equations for excited states in the $\mathfrak{sl}(2)$ sector. They were used in \cite{Gromov:2009zb} to numerically calculate the spectral curve for the Konishi operator up to relatively large values of the coupling constant $\lambda \simeq 700$. The resulting strong coupling expansion takes the following form 
\beq
\DK(\lambda)=2 \lambda^{\frac{1}{4}}
\left(1+\frac{a}{\lambda^{\frac{1}{4}}}+\frac{b}{\lambda^{\frac{1}{2}}} +\ldots \right)\,,
\eeq
with $a^{\textrm{\tiny TBA}}=0.0044$ and $b^{\textrm{\tiny TBA}}=1.0095$. On the other hand, the computations on the string theory side performed in \cite{Roiban:2009aa} suggest different values of the subleading coefficient $b^{\textrm{\tiny string}}=\frac{1}{2}$.  Please note that the leading coefficient has been already suggested in \cite{Gubser:1998bc}.

In this article, assuming $g=\frac{\sqrt\lambda}{4\pi} \gg 1$, we determine $\DK(\lambda)$ to the order $\Op\bigg(\frac{1}{\lambda^{\frac{1}{4}}} \bigg)$ using the asymptotic Bethe ansatz with the BHL/BES dressing factor \cite{Beisert:2006ib, Beisert:2006ez}. We find
\beq \label{konishiabaad}
\DK^{\textrm{\tiny ABA}}(\lambda)=2 \lambda^{\frac{1}{4}}
\left(1+\frac{a^{\textrm{\tiny ABA}}}{\lambda^{\frac{1}{4}}}+\frac{\log{\lambda^{\frac{1}{2}}}}{2 \, \lambda^{\frac{1}{2}}}+\frac{b^{\textrm{\tiny ABA}}}{\lambda^{\frac{1}{2}}} +\ldots \right)\,,
\eeq
with $a^{\textrm{\tiny ABA}}=0$,
$b^{\textrm{\tiny ABA}}=\frac{3}{4}+\frac{U}{4\pi}-\frac{\log \left(1+\pi ^2\right)}{2 }+\log (2)+\frac{\cot ^{-1}(\pi
   )}{\pi }\simeq 0.270086$. Please refer to the next section for the definition of $U$. Interestingly, the result obtained from the asymptotic Bethe ansatz \textit{violates} the analytic structure of the strong coupling expansion advocated in \cite{Roiban:2009aa}. The first two coefficients of the expansion seem to be in agreement with the results of \cite{Gromov:2009zb,Roiban:2009aa} suggesting that the finite size corrections are delayed also for the large values of the coupling constant. It should be stressed that for short operators at strong coupling infinite number of L\"uscher corrections are expected to start to contribute at the same order, in contradistinction to the weak coupling case. Their first manifestation should be the cancelation of the logarithmic term and a finite contribution $b_{\textrm{\tiny wrapping}}=b-b^{\textrm{\tiny ABA}}$. It would be an important test of the $\mathfrak{sl}(2)$ TBA equations \cite{Gromov:2009bc} to prove analytically that this is indeed the case, especially due to the fact that logarithms are often difficult to distinguish numerically.

\section{Strong coupling solution of the asymptotic Bethe equations}
\noindent
In what follows, we will use the $\mathfrak{su}(2)$ asymptotic Bethe equations, although the descendants of the Konishi operator may be found in other simple sectors of the theory. The corresponding state of the $\mathfrak{su}(2)$ spin chain has length $L=4$ and two excitations $M=2$. It can be easily checked that this is the only state with these quantum numbers obeying the momentum constraint. One thus expects that the corresponding solution may be also found from the $\mathfrak{su}(2)$ Bethe
equations at large values of $\lambda$, without the necessity to continue from weak to strong coupling.

For $\sqrt\lambda \gg 4\pi$ it is convenient to use the momentum representation of the Bethe equations. The equations for the Konishi operator are given by
\beq \label{konishieq}
e^{ i\, p\,  L-\log S_{\mathfrak{su}(2)}(p,-p) - 2\, i\, \theta(p,-p)}=1 \,,
\eeq
with the momenta of the individual magnons being $p_1=p=-p_2$. We would like to emphasise once again that by definition this equation is \textit{not} valid at strong coupling for finite value of the length, and thus \textit{not} for $L=4$! Despite this fact, we will attempt to solve it in order to estimate the order of the finite size corrections. The corresponding scaling dimension may be found from \cite{Beisert:2004hm}
\beq \label{delta}
\DK^{\textrm{\tiny ABA}}=\Delta_0 + \gamma^{\textrm{\tiny ABA}}_{\mathcal{K}}(\lambda)= 4+ 2 \left(\sqrt{1+\frac{\lambda}{\pi^2} \sin^2{\frac{p}{2}}}-1\right)\,.
\eeq
The large $\lambda$ expansion of the dressing factor $\theta(p_1,p_2)$ has been proposed in \cite{Beisert:2006ib}
\beq \label{theta1}
\theta(p_1,p_2)=\sum^{\infty}_{n=0} \left(\frac{\sqrt \lambda}{4\pi} \right)^{1-n} \theta_n (p_1, p_2)\,,
\eeq
\beq \label{theta2}
\theta_n(p_1,p_2)=\sum^{\infty}_{r=2} \sum^{\infty}_{s=r+1} c^{(n)}_{r,s} \left(q_r(p_1) q_s(p_2) -q_r (p_2) q_s(p_1) \right)\,.
\eeq
Here, $q_r$ denote the rescaled higher charges and the coefficients $c^{(n)}_{r,s}$ are given by
\beqa \label{theta3}
c^{(n)}_{r,s}&=& \frac{(1-(-1)^{r+s})\zeta(n)}{2 (-2\pi)^n \Gamma(n-1)} (r-1)(s-1)\times \nonumber\\
&&\frac{\Gamma[\frac{1}{2}(s+r+n-3]\Gamma[\frac{1}{2}(s-r+n-1)]}{\Gamma[\frac{1}{2}(s+r-n+1)]\Gamma[\frac{1}{2}(s-r-n+3)]}\,.
\eeqa
It was argued in \cite{Beisert:2006ez} that the strong-coupling expansion of the dressing phase may be obtained, upon resummation, from the one proposed at weak coupling. 
At the leading order equation \eqref{konishieq} is dominated by the AFS dressing phase \cite{Arutyunov:2004vx}
\beq
e^{-2 \,i \, \theta_0 (p,-p)}=e^{-i \,\frac{4 \sqrt \lambda}{\pi} \left( \cos (\frac{p}{2}) \log (\cos(\frac{p}{2}))\right)}=1
\eeq
and can be consistently solved by assuming $p=\frac{p_0}{\lambda^{1/4}}+ \ldots$
\beq
e^{\frac{i\,p^2_0}{2\pi}}=1 \quad \Longrightarrow \quad p_0= 2 \, \pi \,\sqrt{m}\,.
\eeq
For Konishi operator we choose $m=1$. Please refer to \cite{Arutyunov:2004vx} for further discussion of different values of $m$. In the same way, one can proceed to the next order and determine the next coefficient in the momentum expansion
\beq
p=\frac{2\pi}{\lambda^{\frac{1}{4}}}-\frac{2\pi}{\lambda^{\frac{1}{2}}}+\dots\,.
\eeq
Here, again only the AFS part of dressing factor contributes. Interestingly, upon substituting in \eqref{delta} and expanding in large $\lambda$, the canonical dimension is cancelled by the anomalous dimension part, i.e.
\beq
\Delta=2\, \lambda^{\frac{1}{4}}+0+\Op \left(\frac{1}{(\lambda)^{\frac{1}{4}}}\right)\,.
\eeq
This curious mechanism has been already observed for the $\mathfrak{su}(1|1)$ Bethe equations in \cite{Arutyunov:2005hd} and is intrinsic to the properties of the AFS dressing phase. At the next order, however, one finds that all $\theta_n(p_1,p_2), n\geq 0$ contribute. In particular, expanding $\theta_1(p,-p)$ to the leading order in $\lambda$, one finds
\beqa
2\,i\, 
 \theta_1 (p,-p)&=&\frac{i \,\pi \, \log (\lambda )}{\sqrt{\lambda }} -\frac{2 \,i \,\pi  \log
   \left(\frac{1}{4} \left(1+\pi ^2\right)\right)}{\sqrt{\lambda }}\nonumber\\
&&+\frac{4\,
   i\, \cot ^{-1}(\pi )}{\sqrt{\lambda }}+\Op\left(\frac{1}{\lambda^{3/4}}\right)\,.
\eeqa
The appearance of the logarithm of the coupling constant enforces to seek the solution to \eqref{konishieq} in the form
\beq \label{momentum}
p =\frac{2\pi}{\lambda^{\frac{1}{4}}}-\frac{2\pi}{\lambda^{\frac{1}{2}}}+\frac{1}{\lambda^{\frac{3}{4}}} (p_2+p_{2l} \log{\lambda})+\ldots
\eeq
All higher $\theta_n(p,-p), n\geq 2$ depend only on the first term of the above expansion. By expanding first few of them, we have found that the leading contribution can be cast in the following form
\beqa
2\,i\, 
\theta_n (p,-p)&=&-\frac{2\,i\,}{\sqrt{\lambda}} \left[\left(1+\pi ^2\right)^{\frac{3}{2}-n} (2 n-5){!!} \times  \right. \nonumber \\
&& \left.\cos \bigg((3-2 n) \cot ^{-1}(\pi )\bigg) \,
   \zeta (n)\right] \nonumber \\
&&+\Op\left(\frac{1}{\lambda^{3/4}}\right)\,.
\eeqa
Clearly, their overall contribution $\sum^{\infty}_{n=2} 2\,  i\, \theta_n (p,-p)$
is divergent. This sum, however, may be performed by means of the Borel summation. Indeed, since 
\beq
(2n-5)!! =\frac{2^{n-2}\,\Gamma\left(n-\frac{3}{2}\right)}{\sqrt{\pi}}\,,
\eeq
and using the integral representation of the gamma function we find 
\beq \label{result}
\sum^{\infty}_{n=2} 2\, i\, \theta_n (p,-p) =\frac{i\,U}{\sqrt \lambda} + \Op \left(\frac{1}{\lambda^{3/4}} \right)\,,
\eeq
with $U=-0.98230989031886\dots$.
Alternatively, one can use the integral representation of the zeta function, or directly its series representation. All these regularisations lead again to \eqref{result}. We have also used the DHM integral representation \cite{Dorey:2007xn} to numerically calculate the strong coupling expansion of the dressing phase and found agreement with the analytic expressions proposed in this letter. Combining all the terms together, it is straightforward to determine the coefficients $p_2$ and $p_{2l}$ in \eqref{momentum}. Upon substituting \eqref{momentum} in \eqref{delta} and expanding, one finds \eqref{konishiabaad}.\\ 
\begin{acknowledgements}
We would like to thank Lisa Freyhult, Sergey Frolov, Nikolay Gromov, Tomasz \L{}ukowski, Matthias Staudacher and Arkady Tseytlin for discussions. Adam Rej is grateful to STFC for support. Fabian Spill would like to thank the Deutsche Telekom Stiftung for a PhD fellowship. 
\end{acknowledgements}

\end{document}